\documentclass[10pt, conference, compsocconf]{IEEEtran}
\usepackage{graphicx}
\usepackage{color}
\usepackage{tabularx}
\usepackage{epsfig}
\usepackage{listings}
\usepackage{setspace} 
\ifCLASSINFOpdf
\else
\fi
\begin{document}
%
\title{Semantic Support for Log Analysis of Safety-Critical Embedded Systems}


\author{\IEEEauthorblockN{Alessio Venticinque, Nicola Mazzocca}
\IEEEauthorblockA{Dep. of Electric Engineering and Information Technologies\\
University Federico II of Naples\\
Naples, Italy\\
\{alessio.venticinque,nicola.mazzocca\}@unina.it}
\and
\IEEEauthorblockN{Salvatore Venticinque, Massimo Ficco}
\IEEEauthorblockA{Dep. of Industrial and Information Engineering\\
Second University  of Naples\\
Aversa, Italy\\
\{salvatore.venticinque, massimo.ficco\}@unina2.it}
}


%


\maketitle

\begin{abstract}
Testing  is a relevant activity for the development life-cycle of Safety Critical Embedded systems. 
In particular, much effort is spent for analysis and classification of test logs from SCADA subsystems, especially when failures occur. 
The human expertise is needful to understand the reasons of failures, for tracing back the errors, as well as to understand which requirements are affected by  errors and which ones will be affected by  eventual changes in the system design. 
Semantic techniques  and full text search are used to support human experts  for the analysis and classification  of test logs, in order to speedup and improve the diagnosis phase. 
Moreover, retrieval of tests and requirements, which can be related to the current failure, is supported in order to allow  the discovery of available alternatives and solutions for a better and faster investigation of the  problem.
\end{abstract}


%
\IEEEpeerreviewmaketitle

\section{Introduction}
The design and development of Safety-Critical Embedded Systems (SCES) is a relevant activity in many application fields such as railway, automotive, aerospace, health, etc.  
The life cycle of a new product must follow both regulatory constraints and challenging requirements. 
Moreover, it can be affected by several conflicting objectives to be achieved, such as minimization of costs, improvement of performance, and short time to market. 
Both service levels and deadlines of commitments must be satisfied without failing the more and more challenging safety constraints.
Testing represents a relevant phase that both stresses the system and certificates the achievement of the objectives
\cite{FiccoWSRL}. 

Safety-critical real-time systems require thorough Validation and Verification (V$\&$V) activities, often regulated by international standards (e.g., CENELEC \cite{test2} for railway signaling systems), aiming to guarantee the correctness of the system and its compliance to functional and safety requirements. 
In general, the testing activity involves a high number of testing scenarios (usually more than ten thousand tests), and a high number of testing phases following the whole development cycle. 
For a safety-critical system, the number of test cases grows because of the need to
find faults that are not usually discovered by tests only based on functional requirements. 
The high number of test cases generates a huge amount of test log characterized by a large heterogeneity, typical of $big data$. 
Analyzing such big data can be very time critical. 
Many errors may occur during the test  and  a support for their analysis, classification and trace back could leverage and speed up the task of handling those activities which complement the human expertise. 
 
In this paper, in order to improve performance of test analysis and classification, we will explore the experiences of integrating semantic reasoning and text analysis technique on test log in the complex mission critical railway industry. 
A railway ontology has been designed, as well as a dedicated tool has been  developed  to support semantic test analysis. 
The proposed solution has been designed to support the test of SCADA (Supervisory Control and Data Acquisition) of railway system, in a simulated operational environment, when a prototype of the system is not yet available, or cannot be used to reproduce safety critical scenarios.

The rest of this paper is organized as follows. 
Related work is presented in Sec.~\ref{rw}. 
Sec. \ref{testing} describes the testing process of SCES, focusing on the test report analysis phase. 
The proposed approach is presented in Sec.~\ref{approach}. 
Case study is presented in Sec.~\ref{casestudy}. 
Sec.~\ref{tool} describes the framework designed to support semantic analysis of test logs. 
Conclusions and future work are presented in Sec.~\ref{conc}. 
\section{Related work}
\label{rw}
The design and development of SCES must respect  country normative and the not functional requirements in order to mitigate the effect of a failure. 
To achieve this objective the system has to satisfy 
reliability, availability, maintainability and safety (RAMS) criteria. 
For example, the CENELEC normative (EN 50126-50128-50129) is an European directive for the development of a system in the railway domain. 
In order to guarantee the safety of the system, all  dangerous situations must be evaluated in all operational modes and environmental conditions. 
Such target is obtained through the testing of the safety related requirements, which are expressed in natural language.
Log analysis and interpretation is the most critical activity that involves the human expertise in order to trace back the errors and to identify the right corrections to the project, which do not affect other requirements. 
Although, several commercial and open-source log analysis tools are available \cite{Splunk, ArcSight}, usual practice is to manually analyze the log, whereas few organizations invest in developing tools to automate  the analysis process.
Testers use application logs to verify manually the functional conformance of software with the specification.

In the literature, several works investigated techniques for Big Data interpretation \cite{bigdata1}, as well as for log analysis and classification of test data \cite{testanalysis}. 
Semantic techniques for effective retrieval of information have been  adopted \cite{semantic2,CISIS14}. 
Moreover, the exploitation of ontologies to annotate has been proposed for building knowledge to be processed by intelligent reasoners \cite{semantic,semantictest}. 
In this paper, we want to show how to apply these techniques, by means of open-source tools, to test log analysis of a real case study in the railway domain.

Concept-based retrieval methods have attempted to tackle these difficulties by using manually built thesauri, 
by relying on term co-occurrence data, or by extracting latent word relationships and concepts from a corpus. In \cite{Egozi:2011:CIR:1961209.1961211}  a method that augments keyword-based text representation with concept-based features is presented for  generation of new text features automatically. In fact, due to the lack of labeled data, traditional feature selection methods cannot be used.  This is relevant in different application contexts.  In Biomedical research, retrieving documents that match an interesting query is a task performed quite frequently.  In this field, semantic indexing of the results of a query is presented in \cite{Lourenco20103444}.  Relevant terms in a document  emerge from a process of Named Entity Recognition that annotates occurrences of biological terms (e.g., genes or proteins) full-texts. The system is based on a learning process that 
starts from a set of manually classified documents. The resulting network of items implements the semantic indexing of documents and terms, allowing for enhanced navigation and visualization tools, as well as the assessment of relevance for new documents. To solve the limitations of keyword-based models, the idea of conceptual search, is addressed in \cite{Fernandez2011434}  too. This work investigates the definition of an ontology-based IR model, oriented to the exploitation of domain Knowledge Bases to support semantic search capabilities in large document repositories, stressing on the one hand the use of fully fledged ontologies in the semantic-based perspective, and on the other hand the consideration of unstructured content as the target search space. The integrated utilization of semantic techniques and text document processing technologies has been extended also to classification and discovery of services. In \cite{5744076} web service discovery is addressed. Authors propose an ontology framework  for achieving functional level service categorization, given that  a vast majority of web services exist without explicit associated semantic descriptions. The proposed approach involves semantic-based service categorization and semantic enhancement of the service request. 
\section{Testing of Safety-Critical Embedded Systems}
\label{testing}
In the testing process of SCES it is possible to identify the following activities: test definition, test execution, test report analysis and test report document drawing up \cite{alessio1}.
\begin{itemize}
 \item \textit{Test Definition} starts from system requirements, where tests are manually defined and recorded in test cards. Then, the test cards are used for a manual execution or translated into test scripts, in a proprietary formal language, for an automatic execution. It could consume up to 25\% of the total effort that is necessary for testing.
 \item \textit{Test Execution} usually represents the 15\% of the testing effort. In the current practice, where an automatic test execution environment is available, there are interoperability problems due to different proprietary formalisms from heterogeneous providers. 
\item During  \textit{Test Report Analysis} most of the efforts is spent (up to 50\%) because it is manually performed and requires skill and high level human expertise.
\item \textit{Test Report document Drawing Up} depends on the previous results. It is manually performed. The remaining 10\% of the testing effort is spent here.
\end{itemize}
Tests must cover the functional specification of the system, and must be compliant with the related normative. 
Test descriptions provide a set of behaviors of the system, a representation of the state of entities,  signals and anomalies. 
They also include  checks to be verified in order to declare the test passed, also in terms of not functional requirements.
The test plan gives the evidence that all safety and testability requirements are covered, and specifies which set of tests covers each functionality.

\subsection{Test execution and  analysis}
To be noticed that  a safety-critical control system typically manages equipments which are capable to cause damages,
and its malfunction may lead to hazardous situations and possible accidents. 
Therefore, the system must achieve the required reliability the first time it is put into operation. 
This gives few chances to test SCADA of safety critical system in its real operational environment without threat for human lives and
properties \cite{scada2}. 
For these reasons, the testing of such systems is widely based on simulated environments. Another attractive feature of the approach is that, the system can be tested not only in the normal operation of the environment, but also in adverse conditions representing the non-healthy operation of the environment.
Thus, the robustness and tolerance of the system to an unexpected system behavior can be assessed.
Simulated environments are also used to allow the test execution activity as soon as possible in early stages of the development cycle,
when a prototype of the system is not yet available.

On the other hand, the use of simulated testing environments allows to increase the parallelism of the test execution, in fact different replicas of the testing environment can be deployed on different workstations. 
The Cloud paradigm allows also to small and medium enterprises to deploy such replicas when they need, using   the computing power  according to the deadline they need, paying per use. 
In a such elastic and scalable scenario, test execution is not an issue, but the analysis of test log could be a critical activity.
The complexity of test log analysis is related to the fact that, even a SCADA system used in a small installation generates thousands of potentially alarming log entries per day \cite{loganalysis}.
Thus, the size (and high dimensionality) of logs from all simulation of a number of projects make manual inspection practically infeasible.
This is a relevant and challenging problem to tackle.  
It is challenging because, in the past the analysis of system logs has been applied to other domains (e.g., the security domain in \cite{Julisch,Ficco20132}), but failed to deliver convincing results.

\subsection{Big Data  requirements}
Logs that have the characteristics of Volume,Velocity and/or Variety can be defined Big Data \cite{ibm}.
Volume refers to the fact that we are dealing with ever-growing data expanding beyond terabytes into petabytes, and even exabytes (1 million terabytes). In fact, we have to take into account not only the overall logs historically produced by an enterprise, but also all the documentation (requirements, test description, documentations, external repositories) on which the reasoning will be performed. 
Variety refers to the fact that logs characterized by data that often comes from all the testing replicas belonging to different  working teams and from different projects.
Finally, the third characteristic, that is velocity 
that is due to the high number of logs which comes when the deadline is approaching and the high volume of parallel simulations increases. In this case, it is important to perform analytics against the volume and variety of data while it is still in motion, not just after \cite{ibm}. It means to find relationships about logs and the produced by other teams which are working in parallel on other parts of the system.
Big Data is so complex and large that it is really difficult and sometime impossible, to process and analyze them using traditional approaches. In fact, traditional relational database management systems (RDBMS) can not handle big data sets in a cost effective and timely manner. 
The challenge is to find a way to transform logs and other documents into valuable information. To capture value from big data, it is necessary to use technologies and techniques that will help experts to analyze, correlate and retrieve logs, test and requirements.
%
Most of the NoSQL databases, have as the main objective, the achievement of  scalability and higher performance.
Some concepts in NoSQL databases  can be exploited in such a SCADA system \cite{nd}.
 Sharding, also referred to as horizontal scaling or horizontal partitioning,  is a partitioning mechanism in which records are stored on different servers according to some key. 
Consistent hashing 
uses the same hash function for both the object hashing and the node hashing. This is advantageous to both objects and machines. 
MapReduce \cite{mapreduce}  is a programming model for processing and generating large data sets.
It is typically used to do distributed computing on clusters of computers.  Supporting of versioning of datasets in distributed scenarios is another feature that leverages the test management.

\section{The testing workflow}
\label{approach}
We propose a semi-automated approach for log  analysis  that relies on  the integration of  semantic reasoning and text analysis techniques.
Semantic techniques can be used to  resolve ambiguity and improve precision if semantic annotations or any kinds of metadata are available.  On the other hand, a text search engine  is able to extract  all keywords from requirements, which are described in natural language and which have not been explicitly included into metadata.
Furthermore, semantic reasoning can be used to infer  from annotations and keywords, by ontologies, some implicit relationships which are not easy to be recognized.

\begin{figure}[ht]
\centering
\includegraphics[width=3.6in,natwidth=804,natheight=622]{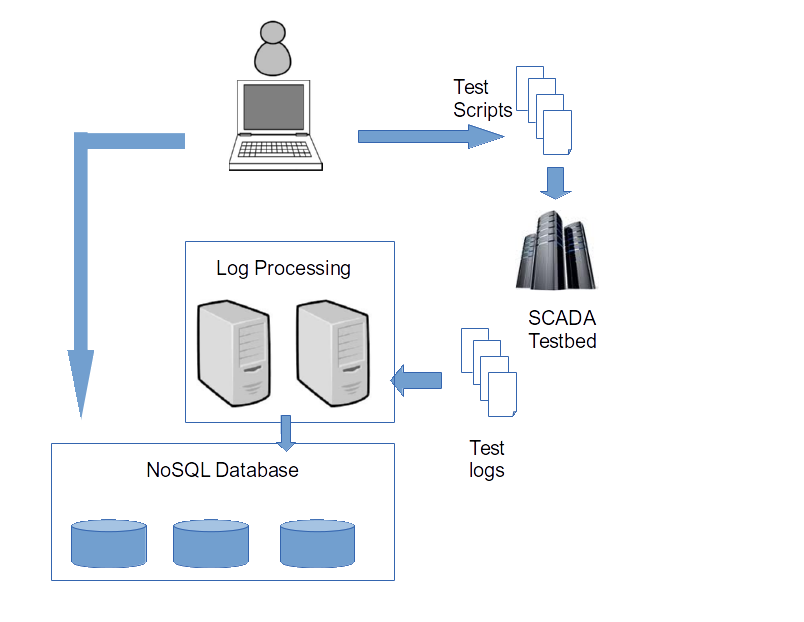} 
\caption{Flow Chart}
\label{arch}
\end{figure}

In Figure \ref{arch}, the architecture of the framework that supports the test execution and analysis is shown.  
The user is able to submit from his workstation tests, which are scheduled by a dispatcher \cite{r.2008-2} over a distributed Cloud infrastructure. 
The logs produced by the completed simulations are processed, classified and stored in a NoSQL storage. When the user is alerted about the availability of new results, he can look for inferred information about the  current tests and the knowledge base of historical documents, eventually retrieving the relevant ones. The NoSQL storage will contain not only the logs, but also test scripts and requirements which have been previously processed and classified.
\vspace{0.8 cm}

\begin{figure}[ht]
\centering
\includegraphics[width=3.4in]{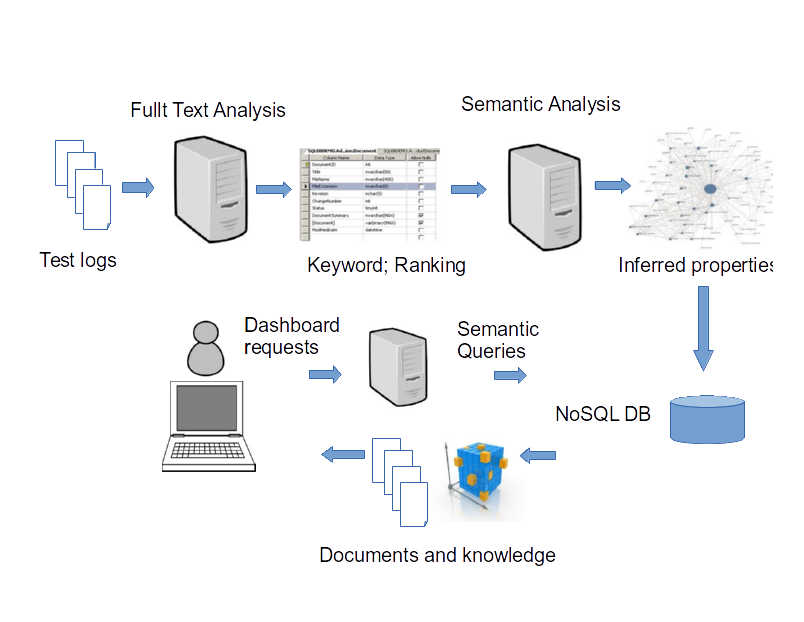}
\caption{Architecture}
\label{workflow}
\end{figure}

In Figure \ref{workflow}, the workflow of the log analysis phase is shown.  On the upper side, of the figure logs are processed by a text-full analyzer. It extracts keywords and their score for each document providing also full-text search capabilities. For each document, the extracted features are used by a semantic engine that infers the relationships between the documents and the knowledge base using a domain ontology.
Such metadata and the document itself are stored in a NoSQL document store.
On the lower side, the user is provided with a web dashboard that allows him to browse the properties of the log of interest and the related documents and concepts over the domain ontology.
The dashboard provides a friendly interface for  querying the NoSQL DB and retrieving the relevant documents. User interactions starts a semantic query over the NoSQL knowledge base. 
\section{ERMTS SCADA}
\label{casestudy}
\begin{figure}[ht]
\centering
\includegraphics[width=3in,natwidth=600,natheight=280]{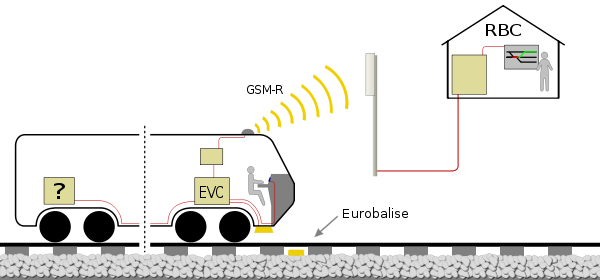} 
\caption{ERMTS Case Study}
\label{ermts}
\end{figure}

 ETCS (European Train Control System) is an ERTMS basic component: it is an Automatic Train Protection system (ATP) to replace the existing national ATP-systems. 
The ERTMS/ETCS  application has three levels that define different uses of ERTMS as a train control system, ranging from track to train communications (level 1) to continuous communications between the train and the Radio Block Centre (level 2). Level 3, which is in a conceptual phase, will further increase ERTMS potential by introducing a moving block technology. This Pilot Application (PA) deals with ERTMS level 2.
ERTMS level 2 is designed as an add-on to or overlays a conventional line already equipped with line side signals and train detection equipment which locates the train. ERTMS level 2 has two main subsystems:
\begin{itemize}
 \item \textit{ground subsystem}: collects and transmits track data (speed limitations, signal-status, etc.) to the on-board subsystem, it is composed by Interlocking and Radio Block Centre;
\item  \textit{on-board subsystem}: analyzes data received from the ground and elaborates a safe speed profile.
\end{itemize}
Communication between the tracks and the train are ensured by dedicated balises (known as Eurobalises) located on the trackside adjacent to the line side signals at required intervals, and a GSM-R connection between the train and the RBC (Fig. \ref{ermts}). The balises contain pre-programmed track data. The train detection equipment sends the position of the train to the control center. The control center, which receives the position of all trains on the line, determines the new movement authority (MA) and sends it to the train. The on-board computer then calculates its speed profile from the movement authority and the next braking point. This information is displayed to the driver. 

\subsection{ERMTS/ETCS SCADA subsystem}
The SCADA subsystem of ERMTS/ETCS handles the regulation and the supervisor of trains movement along the line in a centralized way. It must be compliant with the CENELEC specifications.
Typically such subsystem have functionality for the remote control or automatic control, in order to support, by a single human operator, the management of  different stations from one place,  to track trains and control railway signals, to set automatically routes and align the train ride to timetables; integrating and super visioning the entire network.  Others functionality performed are the monitoring and the diagnosis of the state of the infrastructure through different kind of sensors and subsystems distributed along all the line (passenger information, statistical data collection, etc.). 
In particular, the SCADA subsystem provides tools for automating planning and rooting optimization, monitoring of the traffic on the network in real time  and efficient platform allocation in station. Often the optimization about routing and platforms are static, based on pre-configured data, but in many cases the SCADA subsystem supports operator decision, offering problems resolutions to optimization chooses based on dynamic data. The SCADA subsystem allows also the dispatching of commands, setting restriction on the line based on a particular state that could be unexpected or programmed. 
The SCADA subsystem could receive also alarms from the network (warning lights or acoustic alarms), from  sensors or other subsystems.
The entire network could be divided in more supervisor nodes, with one or more operators, due to the big amount of information presented to the operator and the effort of the management of this activities. In this situation, the SCADA subsystem could have a hierarchical organization, with communication between nodes and messaging services exported to operators. In order to have an idea of the large number of parameters to be monitored and analyzed by such system we can think that for a network of   middle high complexity there are 10000 object that can changes state in real time, just  concerning the binary sections and the trains moving along the line. In addiction to this, consider that the controlled points  could be over 5000. Over 3000 are the signaling object and around 1500 the text string used to communicate. 
%
\subsection{Case study}
A subsystem of the railway infrastructure is the RBC (Radio Block Center). It is connected to TMS (Traffic Management System), to the trains  and to the  IXL (Interlocking). It receives information from the fields and  gives an alarm if  safety conditions are not met. For example, it can alert the operator if the maximum possible number of trains present in line is reached. The RBC is a SCADA  subsystem itself.
For our case study, we consider a requirement of the SoM (Start of Mission) function derived from UNISIG normative. 
The requirements about SoM describe the procedure followed by on board unit (OBU) and wayside equipment in order to start the ride of the train on a ERTMS/ETCS level 2 line. We consider in particular the following requirement: 
\textit{``When the OBU performs a start of mission, it must send a valid SoM Position Report to the RBC. If RBC detects the position of the OBU as valid, it shall send the MA (Movement Authority) to the OBU.
If RBC detects the position of the OBU as invalid, it shall not send the MA to OBU and shall activate the  emergency procedure: the RBC must send an Emergency Brake Order to OBU.''}

We can consider the following scenario to be tested in order to verify the compliance of the system with the  selected requirement: 
\textit{``The train is in shunting mode and positioned before a switch point in Position AB, then train capts the BG A' and move in position Y. Then the train moves in position X and performs the Start of Mission.
During SoM, the ETCS on board sends a  SoM Position Report to the RBC and capts the balise A. The Position Report is invalid because relates to the LRBG A (Last Relevant Balise Group) that is on another track. As the Position Report is invalid, the RBC could consider the train in a wrong place and could deliver a wrong MA. The RBC does not send the MA to the OBU, but it sends an invalid position alarm and the  order to brake immediately.''}

As example for our case of study we present the following test script for the SoM (start of mission) function:

\vspace{0.6 cm}
\footnotesize
\begin{verbatim}
//set the state of all entities in line
//to default
For each entities in network set:
   State[i]= initial_state[i]
//Force the state of  entities
//to the wanted value
For all entities in the location defined:
   State[j]= setStateTo[j]
//Stimulate the system component with signals
For each Input in I
   stimulate Component[i] with Input[i]
//Monitor of the output and checks
For each Output in O
   check Output of O[i] equals to condition C[i]
// stimulate with SoMcommand the OBU
stimulate Train[i] whit Input[“MakeSom”]
//Monitor if SoM is  performed
check OBU send SoM Position Report to RBC
check RBC send MA to OBU
[...]
\end{verbatim}
 \normalsize

In the first part of the script  the state of the system under test is declared. This is a set of values characterizing the railway network and the internal state. 
Other sections of the test script include  an input sequence and an output sequence. The input sequence is a list of stimuli sent to the system, they could be command sent by an operator to HMI subsystem or a change of status forced to simulate an anomaly behavior of the system. The output sequence  is a set of data or control actions which are produced by the system in response to a certain input.
For each state, input sequence and output sequence, some checks are declared  to  monitor  the right execution of the test. The checks operate also as break point, if the result of the test is false, the next sequence (input or output) is not scheduled.
When this test is executed into the simulated environment  it fails because the RBC did not sent the MA to OBU. In particular, the test log produced can be resumed by the following code.

\footnotesize
\begin{verbatim}
//set the state of all entities in line
//to default 
//For each entities in network set:
Time 353 State[i]= initial_state[i]
Time 354 State[i+1]= initial_state[i+1]
[...]
//Force the state of  entities
//to the wanted value
//For all entities in the location defined:
Time 555 State[j]= setStateTo[j]
Time 556 State[j+1]= setStateTo[j+1]
[...]
//Stimulate the system component with signals
//For each Input in I
Time 767stimulate Component[i] with Input[i]
Time 777 stimulate switch point 32
      position C_B        
[...]
//Monitor of the output and checks
//For each Output in O
Time 888 check Output of O[i]
      equals to condition C[i]
Time 889 check position  switch point 32 
      equals to “C_B”  True
[...]
// stimulate with SoM command the OBU
Time 999 stimulate Train with Input[MakeSom]
//Monitor if Som is  performed
Time 1000 check OBU 
     send SoM Position Report to RBC TRUE
Time 1001 check RBC send MA to OBU FALSE
Time 1002 test failed
Test Stopped
(No others operations are executed)
\end{verbatim}
\normalsize

The output of the test execution is a set of logs, the monitor of all variables about simulated subsystems and their output. 
The  log is the output of the script executed in which timestamps  and the value assumed by variables to check  are reported. Therefore, the log contains the state, the input sequence as reported into the original file, but with the timestamps. For what concerning the output sequence and the associated checks,  the timestamp and  the value assumed by the output are reported and, in case of error, the  notification of failure. Note that, if a failure is detected the test is declared not passed and following input or output items are not processed. The report of the  executed script is a simple way to verify the test result and the termination mode, but  an accurate analysis  is essential, by an investigation about what happen in each subsystems, in order to understand the cause of failure.
\vspace{0.5cm}
\section{A semantic support for log analysis}
\label{tool}
In order to support the approach described in Section \ref{approach} we developed a prototype that allows for processing and storing log into a graph database, and   provides  to V\&V engineers a dashboard for:
\begin{itemize}
 \item retrieval of requirements and test which are semantically related to those linked to the selected log;
 \item retrieval of those logs  that report a failure similar to the current one,  in terms of entities involved and system variables value.
\end{itemize}
\subsection{Ontology design}
\begin{figure*}[ht]
\centering
\includegraphics[width=5.2in]{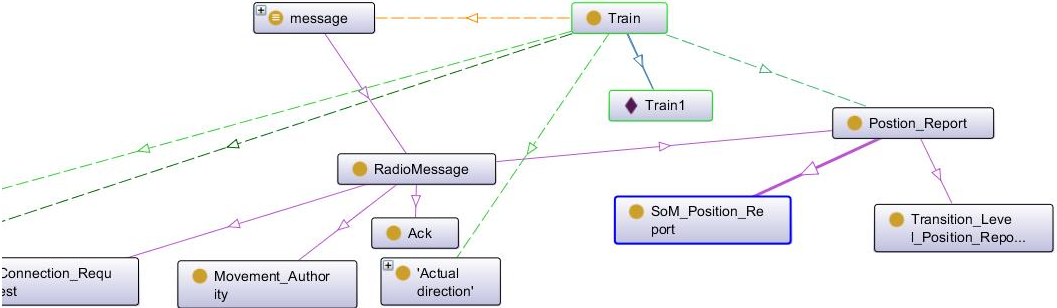}
\caption{Railway Ontology}
\label{ontology1}
\end{figure*}

The railway ontology  represents the knowledge about this specific domain through a set of concepts described in a structured dictionary  consisting of classes, relations, data properties object properties and individuals.  
Classes are concepts that recur into the railway application. They can be both physical and abstract. e.g. the signaling equipment installed on the railway track (balise group), or the distance before a signal by which it needs to react. Relations   describe  logical connection between  two entities of the railway domain. The relations  describe something that  an entity has or something that an entity is. For example, the entity \textit{Train} is linked with property \textit{send} to \textit{Position Report}. It means that a train can send a radio message of type position report. Axioms describe elementary relations, such as the sub-class between concepts or the equivalence. For example, the class \textit{Position Report} is a subclass of \textit{Radio Message}
In the Figure \ref{ontology1}  part of the ontology developed is presented.

The extract shown in Figure \ref{ontology1} focuses on the class \textit{Train}. 
This entity has an individual \textit{Train1}, which is the specific train. Information is communicated through the \textit{radio message} that can be received by the \textit{train} as for example the \textit{MA} or sent as the \textit{Position Report}.
All these information are represented into the picture as a graph where class are drawn as rectangle and relations and axioms between them by edges. Different kinds of edge mean different type of relations. The arch between \textit{Train} and \textit{Position Report} represents a \textit{send} relation. The arch between \textit{Train} and \textit{Train1} means that the train subsystem \textit{has an individual}.
The ontology has been written according to the W3C OWL-RDFS (Ontology Web Language - Resource Description Framework) standard. It is a   knowledge representation language widely used to describe web information. 
\subsection{Document analysis}
For text analysis of requirements, test and logs we  used Apache Solr. $Solr^{TM}$ is a fast open source enterprise search platform from the Apache Lucene$^{TM}$ project. Its major features include powerful full-text search, hit highlighting, faceted search, near real-time indexing, dynamic clustering, database integration, rich document (e.g., Word, PDF) handling, and geospatial search. Solr is written in Java and runs as a standalone full-text search server within a servlet container such as Jetty. Solr uses the Lucene Java search library at its core for full-text indexing and search, and has REST-like HTTP/XML and JSON APIs.
%

$Solr^{TM}$ has been used to extract keyword for cited documents and their score. Such keywords are matched with the domain ontology  to infer RDF triples, which semantically describe each document.
Taking into account the example described in Section \ref{casestudy}, we infer the following metadata from the UNISIG SoM (Start of Mission) requirement.

\footnotesize
\begin{verbatim}
(OBU, perform, SoM)
(OBU, send, SoM Position Report) 
(RBC, send, MA) (OBU, Recive, MA)
(RBC, send, EmergencyBrake)
\end{verbatim}
\normalsize

In the same way, the description of the considered scenario can be described by the following semantic annotation.

\footnotesize
\begin{verbatim}
(OBU, send, SoM Position Report) 
(RBC, send, MA) 
(OBU, Receive, MA)
\end{verbatim}
\normalsize

Of course such information will be linked to the the related document and between them.
Finally, the failed log of our example is processed. Each check instruction of the output sequence from the last control block is  semantically represented. From the following two lines:

\footnotesize
\begin{verbatim}
Time 1000 check OBU1 send SoM 
    Position Report to RBC1 TRUE
Time 1001 check RBC1 send MA to Treno1 FALSE
\end{verbatim}
\normalsize

We infer the following triples:

\footnotesize
\begin{verbatim}
(OBU1, send, SoM Position Report)
(RBC1, receive, Position Report)
(RBC1, send, Ma)
(OBU1, receive, MA)
\end{verbatim}
\normalsize

The log is explicitly linked to the correspondent requirements and scenario in the knowledge base, both to the instance of the ontology and to the text document in the Solr tool.
\subsection{Big Data storage}
 For the implementation of the knowledge base a graph database \cite{graph} has been chosen to improve performance figures as the semantic query will be performed natively. Neo4j \cite{neo}, is an open source, robust (fully ACID) transactional property graph database. Due to its graph data model, Neo4j is highly agile and blazing fast. For connected data operations, Neo4j runs a thousand times faster than relational databases. Nodes store data and edges represent relationships. The data model is called property graph to indicate that edges could have properties. Neo4j provides a REST interface or a Java API. The core engine of Neo4j supports the property graph model. This model can easily be adapted to support the LinkedData RDF model, consisting of Triples. Besides it is possible to add spatial indexes to already located data, and perform spatial operations on the data like searching for data within specified regions or within a specified distance of a point of interest. In addition classes are provided to expose the data to geotools and thereby to geotools enabled applications like geoserver and uDig.
Adoption of Neo4j will  provide custom API's and query languages and many support the W3C's RDF standard, including a SPARQL engine.  This model can easily integrated with the JoWL base front-end. 
\subsection{User dashboard and document retrieval}
For supporting ontology browsing and semantic reasoning we used jOWL.  jOWL is a jQuery plugin for navigation and visualization of OWL-RDFS documents. Moreover, it provides APIs for SPARQL\_DL query.
SPARQL is a language for  semantic querying of a knowledge base. 
\begin{figure}[h]
\centering
\includegraphics[width=3.4in,natwidth=1080,natheight=526]{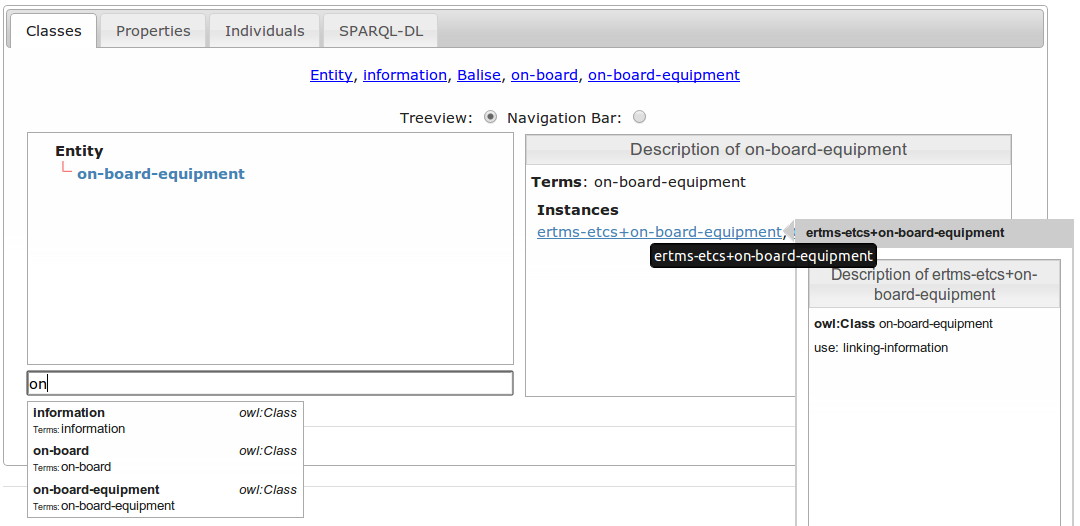} 
\caption{Ontology browser}
\label{jtree}
\end{figure}

We extended its dashboard to support both keyword based search and semantic search.
The web dashboard allows for composing a semantic query starting from the property of a document, a test or a requirement, with concepts and relationships from the ontology.
Such tool can be also used during the  test definition phase in order to retrieve all documents of interest starting from requirements.
During the log analysis, when a failed log has been detected, the  user can look for similar failures in the knowledge base in order to understand the errors.
The user can use the tuples from  failed log  shown in the previous example to search  which test are failed due to checks on the same variables of the same entities. 

One among available documents, resulting from the query that uses the tuples describing the log presented before, contains the following lines: 

\footnotesize
\begin{verbatim}
Time 332 check Linked balise group list
     contains  ETCS5233 FALSE
Time 333 check OBU1 send SoM 
     Position Report to RBC1 TRUE
Time 334 check RBC1 send MA to Treno1 FALSE
\end{verbatim}
\normalsize

represented by the following tuples:

\footnotesize
\begin{verbatim}
(Linked Balise group list, contain, ETCS5233)
(OBU1, send, SoM Position report)
(RBC1, send, MA)
\end{verbatim}

\normalsize
The found test log has  an additional check if it is compared to our example. Retrieving  the correspondent test description is easier to understand how to find the cause of the failure. In this particular case the additional check allows to detect that the position report is sent from an invalid ETCS.

\section{Conclusion}
\label{conc}
We discussed about the critical issues that affect the testing of Safety Critical Embedded systems during  their development. In particular we focused on the test analysis of logs from a SCADA subsystem using semantic techniques and full text search. 
We presented a framework for management of requirements, test scripts and logs that are produced during the simulation of the system for the validation and verification activities.
The framework  uses a graph databased for storing documents and their  properties. It provides a web dashboard by which the user is able to browse the knowledge base starting from a domain ontology or interactive controls that execute semantic and keyword based queries. A real case of SCADA testing, belonging to the railway domain, has been presented. 

\section*{Acknowledgment}
\small
This work has been partially supported by the MIUR under Project PON02\_00485\_3487784 ``DISPLAY'' of the public-private laboratory ``COSMIC'' (PON02\_00669), and by EU with the project CRYSTAL (Critical System Engineering Acceleration), funded from the ARTEMIS Joint Undertaking under grant agreement no. 332830.

\bibliographystyle{IEEEtran}
\small
\begin{spacing}{0.92}
\bibliography{mybib}
\end{spacing}
\normalsize
%
%
%
%
%

\end{document}